\documentclass[%
reprint,
superscriptaddress,
showpacs,preprintnumbers,
nofootinbib,
amsmath,amssymb,
aps,
prx,
showkeys
]{revtex4-1}

\usepackage{graphicx}
\usepackage{dcolumn}
\usepackage{bm}

\usepackage{amssymb}
\usepackage{amsmath}
\usepackage{amsfonts}
\usepackage{xspace}
\usepackage{bm,bbm}
\usepackage{relsize}
\usepackage{siunitx}
\usepackage[usenames,dvipsnames]{color}
\usepackage{float}

\renewcommand{\vec}[1]{\ensuremath{\bm{#1}}}

\newcommand{\ie}{\emph{i.e.}\xspace~}

\renewcommand{\eqref}[1]{Eq.~\ref{eq:#1}}

\usepackage{mathtools}

\DeclarePairedDelimiterX\braket[2]{\langle}{\rangle}{#1 \delimsize\vert #2}

\begin{document}

\title{Topological Bulk Lasing Modes Using an Imaginary Gauge Field} 

\author{Stephan Wong}
\affiliation{School of Physics and Astronomy, Cardiff University, Cardiff CF24 3AA, UK}
\author{Sang Soon Oh}
\email[Email: ]{ohs2@cardiff.ac.uk}
\affiliation{School of Physics and Astronomy, Cardiff University, Cardiff CF24 3AA, UK}

\date{\today}

\begin{abstract}
Topological edge modes, which are robust against disorders, have been used to enhance the spatial stability of lasers.
Recently, it was revealed that topological lasers can be further stabilized using a novel topological phase in non-Hermitian photonic topological insulators. 
Here we propose a procedure to realize topologically protected modes extended over a $d$-dimensional bulk by introducing an imaginary gauge field. 
This generalizes the idea of zero-energy extended modes in the one-dimensional Su-Schrieffer-Heeger lattice into higher dimensional lattices allowing a $d$-dimensional bulky mode that is topologically protected.
Furthermore, we numerically demonstrate that the topological bulk lasing mode can achieve high temporal stability superior to topological edge mode lasers.  
In the exemplified topological extended mode in the kagome lattice, we show that large regions of stability exist in its parameter space.
\end{abstract}


\maketitle

\section{Introduction}

In an attempt of ultimate control of the flow of light, photonic topological insulators (PTIs)~\cite{Ozawa2019} have enabled exciting devices such as unidirectional waveguides and topological lasers that are robust against perturbations and defects.
In particular, the realization of robust topological optical systems has drawn attention for advanced photonics by reducing propagation loss in optical devices~\cite{Khanikaev2012, Wong2020}, for example, quantum computers~\cite{Rechtsman2016, Mittal2016}, photonic neural networks~\cite{Hughes2018} and near-zero epsilon 
devices~\cite{Liberal2017, Huang2011, Saba2017}.

Recently, considerable effort has been made to study non-Hermitian PTIs by engaging topological edge modes to enable a lasing regime with optical non-linearity~\cite{Harari2018, Parto2018}, distribution of gain and loss~\cite{Schomerus2013, Zhao2018, Takata2018, Gong2020} or non-reciprocal couplings~\cite{Bahari2017, Shao2020, Longhi2018}.
For example, the one-dimensional (1D) Su-Schrieffer-Heeger (SSH) model~\cite{Su1979} has been utilized to generate edge states with gain/loss and implement topological lasing devices~\cite{Schomerus2013, Parto2018, Zhao2018}.
A cavity made of topologically distinct PTIs has been proposed to enhance the lasing efficiency by using unidirectional topologically protected edge modes~\cite{Bahari2017, Gong2020}. 

However, the edge-mode-based topological lasers are not appropriate for high power lasers due to the localized nature of the edge modes. 
As an alternative, topological bulk lasers have been proposed to achieve broad-area emission by using extended topological modes based solely on the parity symmetry at the $\Gamma$-point in a two-dimensional (2D) hexagonal cavity~\cite{Shao2020} or by using an imaginary gauge field in a 1D $\mathcal{PT}$-symmetric SSH lattice to delocalize the zero-energy boundary mode over the 1D-bulk~\cite{Longhi2018}. 

Since temporal instability can deteriorate the performance of lasing devices, it is important to study the dynamics and the temporal stability of the topological lasing modes~\cite{Bittner2018}.
Indeed, although the spatial stability of the topological lasing mode is guaranteed based on topological band theory, its temporal behaviour is not necessarily stable due to the non-linear nature of the laser rate equation~\cite{Longhi2018a}. 

In this work, we generalize the topological extended mode on the 1D SSH lattice to higher dimensional lattices. In particular, we demonstrate a topological extended mode on a 2D bulk by using a kagome lattice with a rhombus geometry and an imaginary gauge field.
The topological bulk laser is studied with a gain and loss configuration similar to the $\mathcal{PT}$-symmetric configurations.
We show that the topological extended mode lases and has large stable regions in its parameter space.
We thus demonstrate that a phase-locked broad-area topological lasers can be realized in a 2D kagome lattice.

The structure of this manuscript is as follows: in section~\ref{section:ssh_nh}, we recall the result in Ref.~\cite{Longhi2018} where a topological extended mode is achieved using the 1D SSH lattice~\cite{Su1979} along with an imaginary gauge field. 
Then section~\ref{section:kagome_nh} generalizes this result to higher dimensions. 
An explicit example is carried out on the kagome lattice with rhombus geometry.
Section~\ref{section:kagome_nh_lasing} is dedicated to the study of the non-Hermitian kagome lattice in the active setting where the temporal dynamics of the topological protected mode is studied.

\section{Extended topological mode in 1D lattice}
\label{section:ssh_nh}

Here, we briefly recall the procedure used to delocalize the topologically protected (zero-energy) mode in the SSH lattice, as presented in Ref.~\cite{Longhi2018}. 

\begin{figure}
\center
\includegraphics[width=\columnwidth]{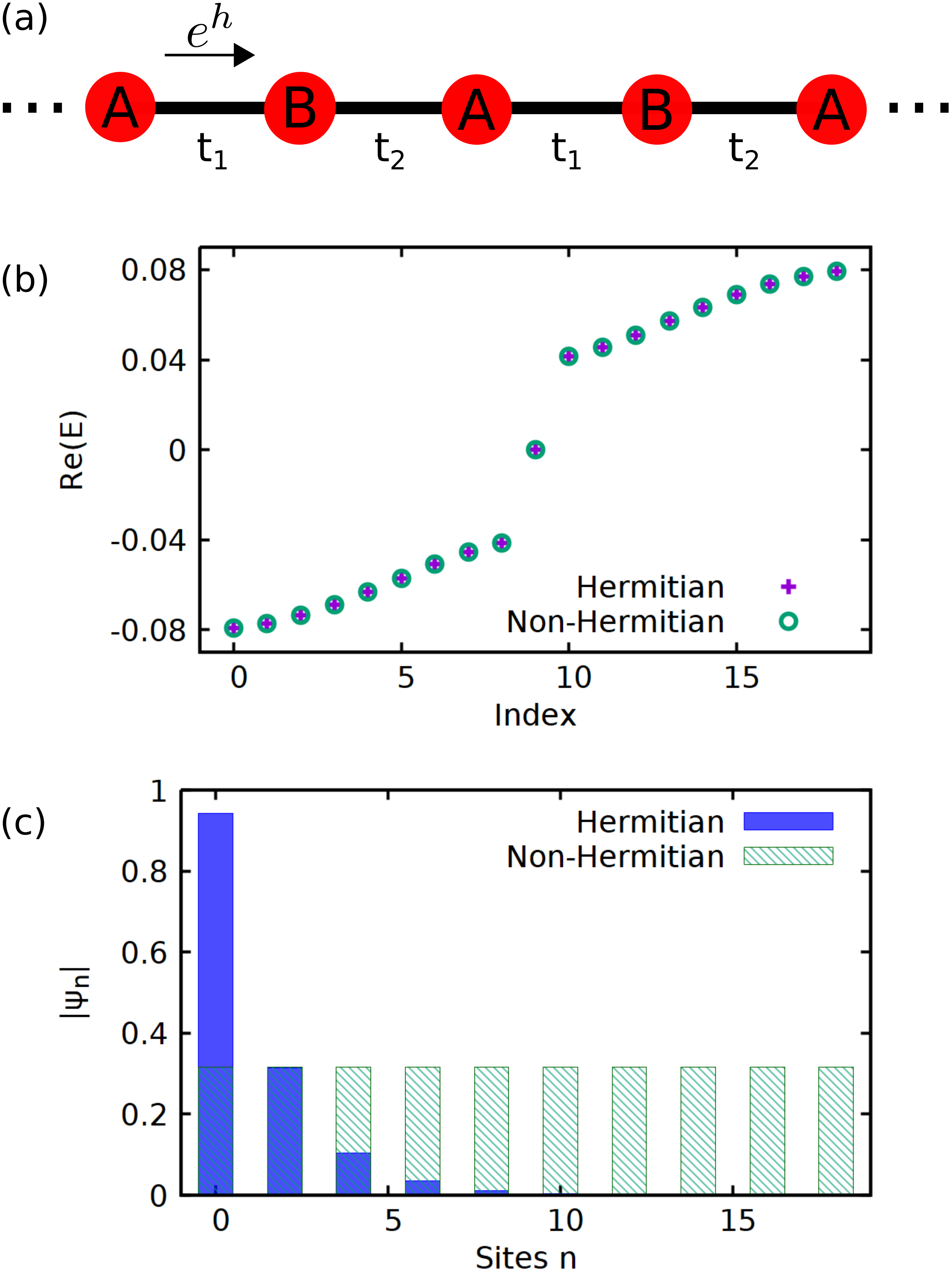}
\caption{
(a) Schematic of a non-Hermitian SSH lattice made of an array of $N_s$ sites. The usual Hermitian SSH lattice corresponds to the case where $h = 0$.
(b) Spectrum of the finite-size SSH lattice in the Hermitian and non-Hermitian setting. The lattice starting from a site A and terminating at a site A. 
(c) Normalized field profile $|\psi_n|$ of the zero-energy mode from the finite-size SSH lattice in the Hermitian and non-Hermitian setting. 
The parameter are chosen such that $N_s = 19$, $E_A = E_B = 0$, $t_1 = 0.02$, $t_2 = 0.06$, and $h = h_0$.
}
\label{fig:ssh_nh}
\end{figure}

The SSH lattice, shown in Fig.~\ref{fig:ssh_nh}(a), is a 1D lattice made of an array of $N_s$ sites. It has a unit cell composed of two sites ($A$ and $B$) and the lattice is characterized by alternating intra- and inter-unit cell couplings given by the real scalars $t_1$ and $t_2$, respectively.

In the non-Hermitian configurations, where an extended mode has been proposed~\cite{Hatano1996, Longhi2018}, an imaginary gauge potential, $\vec{\mathcal{A}} = -i h \vec{e_1}$, is introduced.
In the presence of gauge field, the Peierl's phase modifies the hopping terms by a factor: 
$e^{i \int \vec{\mathcal{A}} \cdot \vec{dl}}$,
with $\vec{dl} = dx_i \vec{e_i}$, the direction of the hopping.
The couplings are therefore modified and become asymmetric. 
Note that the usual pre-factors are here absorbed in $\vec{\mathcal{A}}$, or equivalently in $h$.
The coupling constants get a $e^h$ factor term when hopping from left to right, and a $e^{-h}$ factor term when hopping from right to left.
The coupled-mode equations are then written as follow:
\begin{equation}
\label{eq:rate_equation_ssh_nh_A}
i \frac{d a_n}{dt} = E_A a_n + t_1 e^{-h} b_n + t_2 e^h b_{n-1}
,
\end{equation}
\begin{equation}
\label{eq:rate_equation_ssh_nh_B}
i \frac{d b_n}{dt} = E_B b_n + t_1 e^h a_n + t_2 e^{-h} a_{n+1}
,
\end{equation}
with $a_n$ and $b_n$ the modal amplitudes on the A and B sites at the $n$-th unit cell, respectively. $E_\sigma$ is the on-site energy on the site $\sigma$.

For a finite system, the introduction of the imaginary gauge field will not affect the spectrum~\cite{Hatano1996}. 
This is made explicit here because the system of equations above can be solved with a suitable gauge transformation:
\begin{equation}
a_n = e^{2hn} \tilde{a}_n
,
\end{equation}
\begin{equation}
b_n = e^{2hn} e^{-2h} \tilde{b}_n
.
\end{equation}
where $a_n$ and $b_n$ are solutions of the coupled-mode equations if $\tilde{a}_n$ and $\tilde{b}_n$ are solutions of the Hermitian SSH coupled-mode equations, namely when $h = 0$.

For a SSH lattice starting and terminating on an A site, it is known that the zero-energy mode of the Hermitian SSH lattice reads:
\begin{equation}
\tilde{a}_n = \tilde{r}^n \tilde{a}_0
,
\end{equation}
\begin{equation}
\tilde{b}_n = 0, \; \forall n
,
\end{equation}
where $\tilde{r} = - \frac{t_1}{t_2}$ defined as $\tilde{a}_{n+1} = \tilde{r} \tilde{a}_n$ and satisfies the destructive interference condition on the B sites $t_1 + \tilde{r} t_2 = 0$.

The solution for the non-Hermitian SSH lattice is then written as:
\begin{equation}
a_n = r^n a_0
,
\end{equation}
\begin{equation}
b_n = 0, \; \forall n
,
\end{equation}
where $r = - \frac{t_1}{t_2} e^{2h}$ defined as $a_{n+1} = r a_n$ and satisfies the destructive interference condition on the B sites $t_1 e^h + r t_2 e^{-h} = 0$.

The main effect of this imaginary gauge field is to change the localization property of the modes without affecting the spectrum. 
In particular, one can delocalize the topological protected edge mode over the whole 1D bulk, while keeping its topological protection from the chiral symmetry of the Hermitian topologically protected (zero-energy) mode.
The exponentially increasing or decaying factor is removed by appropriately choosing the gauge field $h$:
\begin{equation}
\label{eq:ssh_nh_h}
h = h_0 := - \frac{1}{2} \ln \left( \frac{t_1}{t_2} \right)
.
\end{equation}

Figure~\ref{fig:ssh_nh}(b) shows that the spectrum of the finite-size Hermitian and non-Hermitian SSH lattice are indeed identical. 
However, Fig.~\ref{fig:ssh_nh}(c) shows, for $t_1 < t_2$, that the field profile $|\psi_n|$ of the zero-energy mode, which is localized on the left edge for the Hermitian case, is extended over the 1D bulk for the non-Hermitian case.

Finally, it is worth noting that although the zero-energy mode is topologically protected, its localization property depends on the coupling constants, and is therefore sensitive to their perturbations. However, for reasonably small perturbations, \ie small enough so that the band gap does not close, the delocalization is not destroyed: the amplitudes remain of the same order of magnitude over the bulk but are simply not equal anymore.

\section{Extended topological mode in \lowercase{d}-dimensional lattice: example on the kagome lattice}
\label{section:kagome_nh}

We now generalize this notion of delocalized (or extended) topological mode over a whole $d$-dimensional bulk. 

\subsection{General framework}

The strategy follows the previous section, namely to find an exact solution of the (topologically protected) boundary state, then use non-Hermiticity to change the localization property of the chosen mode.

\begin{figure}
\center
\includegraphics[width=\columnwidth]{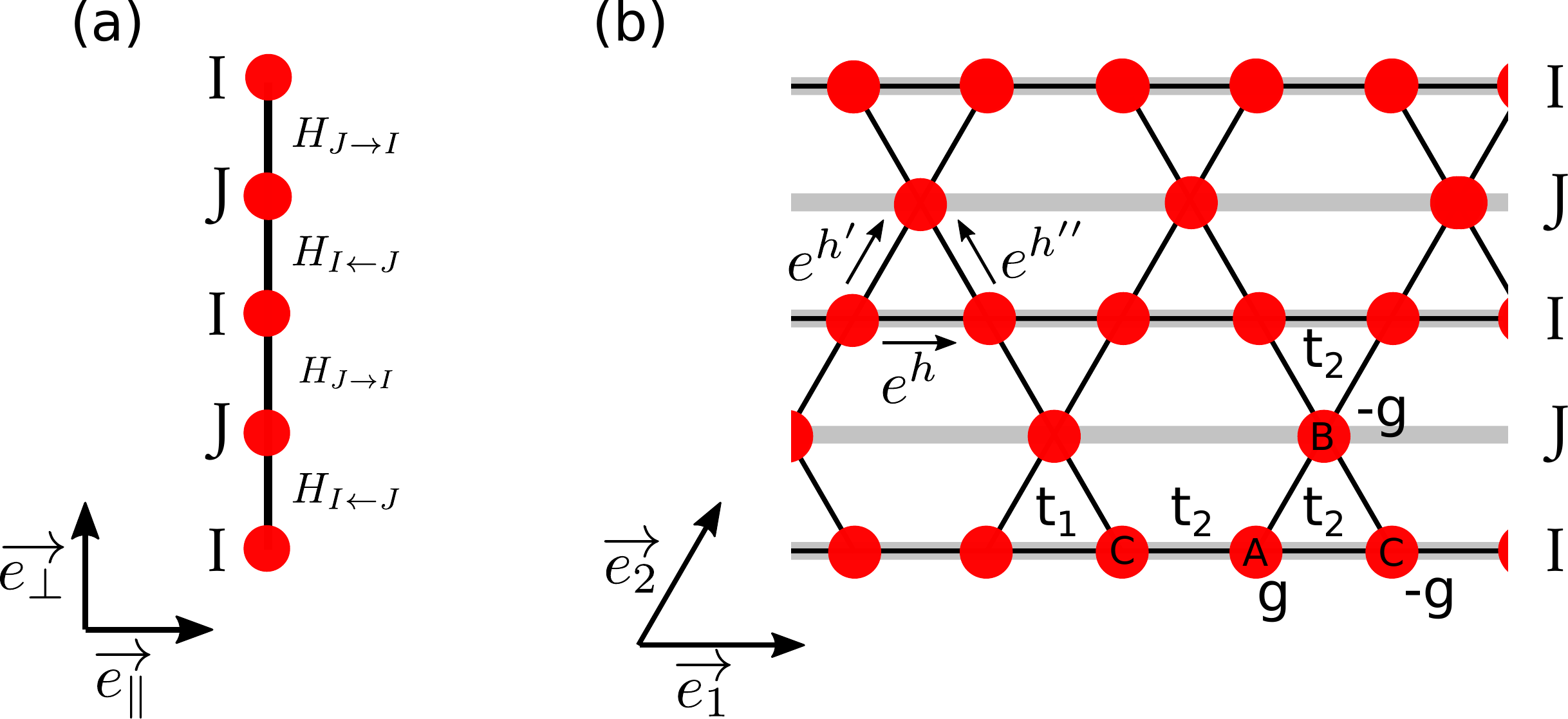}
\caption{
(a) Schematic of a $d$-dimensional lattice in a quasi-1D lattice made of an array of $N_s$ $(d-1)$-dimensional lattices.
(b) Schematic of the kagome lattice draw in the quasi-1D lattice formalism. $e^h$, $e^{h'}$, $e^{h''}$ correspond to the imaginary gauge field introduced for delocalising the topological mode in the non-Hermitian kagome lattice. 
}
\label{fig:quasi_1D_kagome}
\end{figure}

In order to find an exact solution, the procedure follows Ref.~\cite{Kunst2017, Kunst2019}.
One needs to consider a $d$-dimensional lattice as a stack of $(d-1)$-dimensional lattices ($e_{\parallel}$-direction) and with open boundary condition~(OBC) in the remaining $1$-dimensional ($e_{\perp}$-direction) boundary~\cite{Kunst2017, Kunst2019, Mao2010} (see Fig.~\ref{fig:quasi_1D_kagome} for the example of the kagome lattice).
The lattice can then be considered as a quasi-1D lattice with the unit cell composed of two lattice-sites, $I$ and $J$, except that here the lattice-sites represent $(d-1)$-dimensional lattices, as shown in Fig.~\ref{fig:quasi_1D_kagome}.
Additional conditions are assumed such as that the quasi-1D lattice needs to start and terminate on the same lattice-site, and here for our purpose for having a topological protected mode, we forbid direct hopping between the $I$ lattice-sites~\cite{Kunst2017, Kunst2019}.
Therefore, these lattices naturally support the exact disappearance of the wave-function amplitude of n wave functions on the $J$ lattice-sites with n the number of degrees of freedom on the $I$ lattice-site.

From the quasi-1D formalism, the coupled-mode equation is conveniently written as:
\begin{equation}
\label{eq:schrodinger_eq_kagome}
i \frac{d\Psi}{dt} = H_{\text{lattice}} \Psi
\end{equation}
where $\Psi = (\psi_{I,0}, \psi_{J,0}, \ldots, \psi_{I,N})^T$ with $\psi_{I,n}$ and $\psi_{J,n}$ being the modal amplitudes on the $I$ and $J$ lattice-sites in the $n$-th stacked unit cell, respectively. $N$ is the index of the last unit cell. 

The Hamiltonian of the lattice reads:
\begin{equation}
H_{\text{lattice}} = 
\left(
\begin{array}{cccc}
H_I & H_{I \leftarrow J} & 0 & \cdots \\ 
\tilde{H}^\dagger_{I \leftarrow J} & H_J & \tilde{H}^\dagger_{J \rightarrow I} & \cdots \\ 
0 & H_{J \rightarrow I} & H_I & \cdots \\ 
\vdots & \vdots & \vdots & \ddots 
\end{array} 
\right)
\end{equation}
with $H_I$ and $H_J$ being the Hamiltonian of the lattice $I$ and $J$, respectively. $H_{I \leftarrow J}$ and $H_{J \rightarrow I}$ are, respectively, the intra- and inter-unit cell couplings between the $I$ and $J$ lattices. For the Hermitian case, $\tilde{H}^\dagger_{I \leftarrow J} = H^\dagger_{I \leftarrow J}$ and $\tilde{H}^\dagger_{J \rightarrow I} = H^\dagger_{J \rightarrow I}$

For the general $d$-dimensional lattice, the eigenvalue problem $H_{\text{lattice}} \Psi = E \Psi$ yields, for $n = 1, \ldots, N$:
\begin{equation}
\label{eq:coupled_mode_equation_lattice_I_zero}
H_{I,k} \psi_{I,n} + H_{I \leftarrow J} \psi_{J,n} + H_{J \rightarrow I} \psi_{J,n+1} = E \psi_{I,n}
,
\end{equation}
\begin{equation}
\label{eq:coupled_mode_equation_lattice_J_zero}
H_{J,k} \psi_{J,n} + H^\dagger_{I \leftarrow J} \psi_{I,n} + H^\dagger_{J \rightarrow I} \psi_{I,n+1} = E \psi_{J,n}
.
\end{equation}

The condition for destructive interference on the $J$-lattices, $\psi_{J,n} = 0$, is given by:
\begin{equation}
H^\dagger_{I \leftarrow J} \psi^{(i)}_{I,n} + H^\dagger_{J \rightarrow I} \psi^{(i)}_{I,n+1} = 0
.
\end{equation}
From Eq.~\ref{eq:coupled_mode_equation_lattice_I_zero}, the solution with vanishing amplitude on the $J$-lattice therefore gives the additional condition:
\begin{equation}
H_I \psi_{I,n} = E \psi_{I,n}
,
\end{equation}
namely $\psi_{I,n}$ must be an eigenmode of the Hamiltonian on the lattice $I$, labelled $\psi^{(i)}_{I,n}$, with corresponding energy $E = E^{(i)}_I$.

Since we are looking at edge states, one can ask for solutions which exponentially decay or increase, or equivalently solutions that satisfy~\cite{Kunst2017, Mao2010}:
\begin{equation}
\label{eq:ansatz}
\psi^{(i)}_{I,n+1} = r_i \psi^{(i)}_{I,n}
\end{equation}
with $r_i$ being a scalar term representing the decaying amplitudes of the mode inside the quasi-1D lattice.

The solution of the Hamiltonian $H_{\text{lattice}}$ with energy $E^{(i)}_I$ is therefore of the form:
\begin{equation}
\psi^{(i)}_{I,n} = r_i^n \psi^{(i)}_{I,0}
\end{equation}
with $r_i$ satisfying the destructive interference condition on the $J$ lattice:
\begin{equation}
\label{eq:interferences_cond_J_sites}
H^\dagger_{I \leftarrow J} + r_i H^\dagger_{J \rightarrow I} = 0
\end{equation}

In summary, the modes with eigenenergy $E = E^{(i)}_I$, such that $H_I \psi^{(i)}_I = E_I \psi^{(i)}_I$, are the modes which are exponentially localized on one edge and with non-vanishing amplitudes only on the lattice-sites $I$, with adjacent $\pi$-phase difference and with mode distribution corresponding to $\psi^{(i)}_I$ on the lattice-site $I$.

The delocalization of the edge modes is realized by introducing an imaginary gauge field such that $|r_i| = 1$. The non-Hermiticity allows the change of the localization properties while keeping the spectrum unchanged. 

\subsection{Extended topological mode in 2D kagome lattice}

As a concrete example we will look at the case of the kagome lattice as shown in Fig.~\ref{fig:quasi_1D_kagome}.
The kagome lattice is characterized by unit-cells composed of three sites A, B, and C and the coupling strengths between sites are different for intra-unit cell ($t_1$) and inter-unit cell ($t_2$) couplings.

\begin{figure}
\center
\includegraphics[width=\columnwidth]{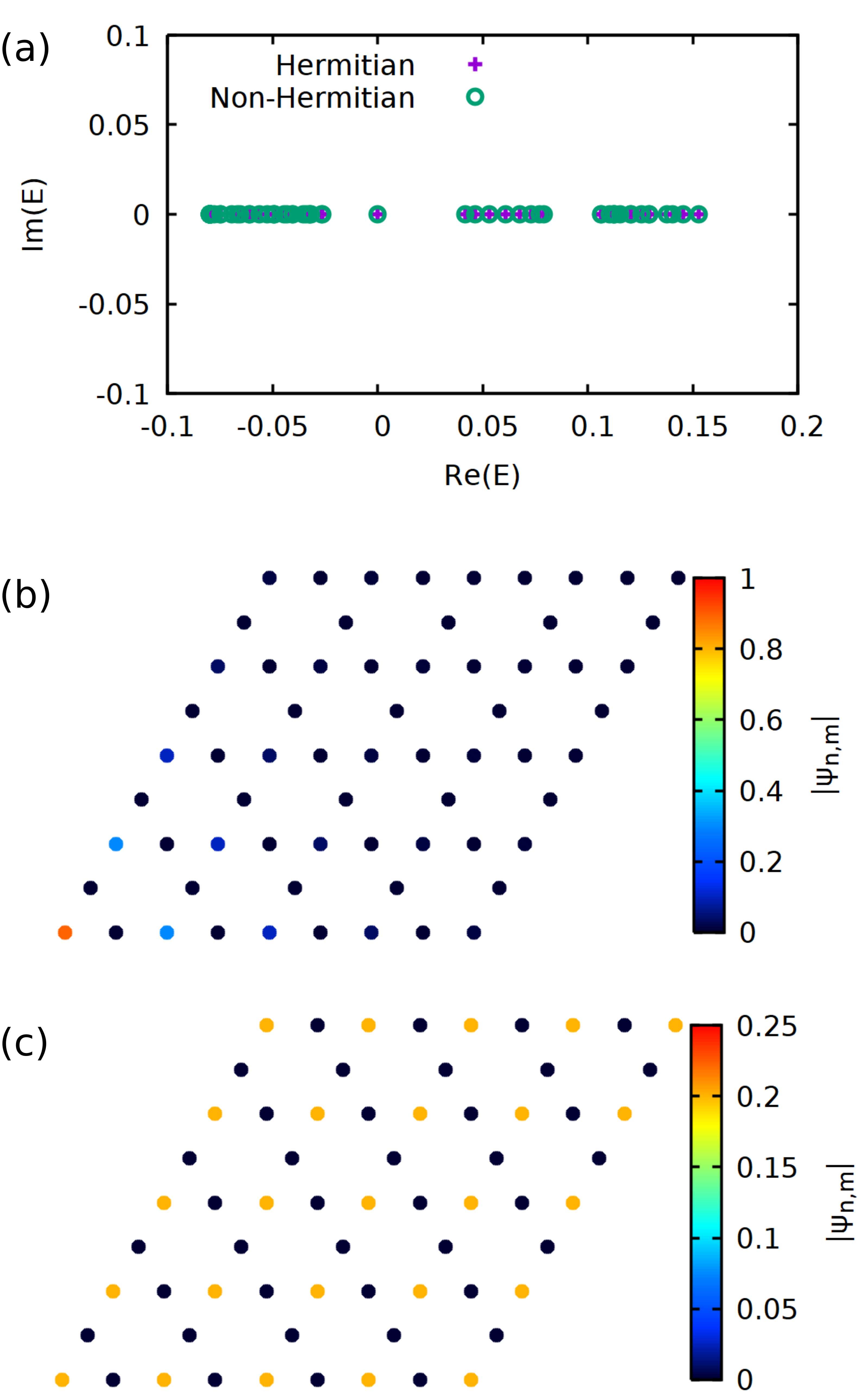}
\caption{
(a) Spectrum of the kagome lattice in the rhombus geometry in the Hermitian and non-Hermitian setting.
The normalized field profile $|\psi_{n,m}|$ of the zero-energy mode of the kagome lattice in the rhombus geometry is plotted in (b) for the Hermitian case, and in (c) for the non-Hermitian case.
Here, there are $N_s = 11$ sites both in the $I$ lattice and the quasi-1D lattice. $t_1 = 0.02$, $t_2 = 0.06$. The Hermitian setting corresponds to the case $h = h' = h'' = 0$, whereas the non-Hermitian setting is for $h = h' = h_0$, $h'' = 0$.
}
\label{fig:kagome_nh}
\end{figure}

Applying the previous results to the kagome lattice in the rhombus geometry (see Fig.~\ref{fig:kagome_nh}(b) for the geometry of the lattice), we have $H_I = H_{\text{SSH}}$, $H_J = E_B$. 
$\tilde{H}^\dagger_{I \leftarrow J} = H^\dagger_{I \leftarrow J} = (t_1, t_1, \ldots, t_1, t_1)^T$ and $\tilde{H}^\dagger_{J \rightarrow I} = H^\dagger_{J \rightarrow I} = (t_2, t_2, \ldots, t_2, t_2)^T$ are ($N_s \times 1$)-matrices corresponding to the intra- and inter-stacked lattice couplings, respectively. $N_s$ is the number of sites on the $I$ lattices. $E_B$ is the on-site energy at the B sites.

The rhombus geometry is interesting since in the $e_{\parallel}$ direction, the $I$ lattices, which are equivalent to the SSH lattice, start with and are terminated by  the same site (here site A). 
In this configuration, the chiral symmetry of the SSH lattice guarantees the presence of the zero-energy mode.
Therefore a boundary state of the kagome lattice with eigenenergy $E^{(i)}_I = E^{(0)}_I = 0$ is topologically protected by virtue of the chiral symmetry in the SSH lattice (lattice $I$).
The corresponding zero-energy mode of this kagome lattice is written as:
\begin{equation}
\psi^{(0)}_{I,m} = r_{0,2}^m \psi^{(0)}_{I,0}
\end{equation}
with $\left[ \psi^{(0)}_{I,0} \right]_n = r_{0,1}^n a_{0,0}$ being the $n$-th component of the zero-energy mode, $\psi^{(0)}_{I,0}$, of the SSH lattice where the interference conditions (Eq.~\ref{eq:interferences_cond_J_sites}) give $r_{0,1} = - \frac{t_1}{t_2}$, and $r_{0,2} = - \frac{t_1}{t_2}$. $a_{n,m}$ is the modal amplitude on the A site at the $n$-th unit cell in the  $m$-th lattice $I$.

Choosing different intra- and inter-unit cell coupling constants,  $t_1 < t_2$ or $t_1 > t_2$, the zero-mode is then exponentially localized, respectively, on the bottom-left or upper-right edge of the SSH lattice with vanishing amplitudes on the B and C sites: it is a (topological) corner mode~\cite{Ezawa2018, Kunst2018, ElHassan2019}. 

Figure~\ref{fig:kagome_nh}(a) shows the spectrum of the kagome lattice in the rhombus geometry and demonstrates the existence of the zero-energy mode, as explained previously. Figure~\ref{fig:kagome_nh}(b) plots the normalized field distribution $|\psi_{n,m}|$ of the zero-energy mode, for $t_1 < t_2$. This shows that the mode is indeed localized on the bottom-left corner.

\begin{figure}
\center
\includegraphics[width=\columnwidth]{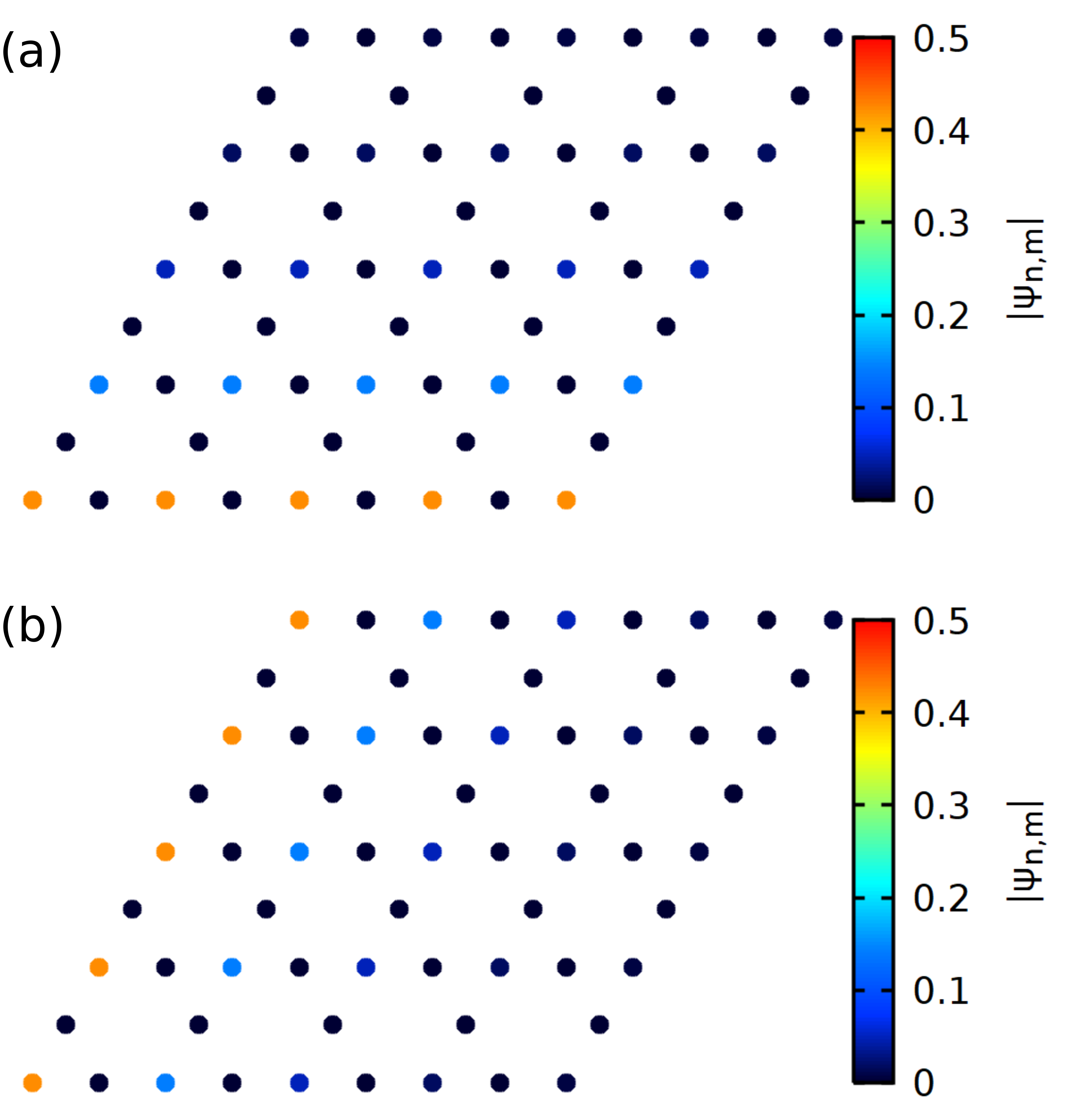}
\caption{
Normalized field profile $|\psi_{n,m}|$ of the zero-energy mode of the kagome lattice in the rhombus geometry with (a) $h = h_0$, $h' = h'' = 0$ and (b) $h' = h_0$, $h = h'' = 0$.
The other parameters are the same as in Fig.~\ref{fig:kagome_nh}.
}
\label{fig:kagome_nh_prelim}
\end{figure}

We use an imaginary gauge field to change the localization property of this corner mode~\cite{Longhi2018, Hatano1996}.
Figure~\ref{fig:quasi_1D_kagome}(b) sketches the gauge potential considered where $e^h$, $e^{h'}$ and $e^{h''}$ represent the phase factors in the couplings between the sites A and B, A and C, and B and C, respectively.
Conditions on $h$, $h'$ and $h''$ are imposed given an imaginary gauge field $\vec{\mathcal{A}} = (\mathcal{A}_1, \mathcal{A}_2)$.
From the Peierls' phase corresponding to $e^h$ we have $\mathcal{A}_1 = -i h$, with $\vec{dl} = \vec{e_1}$. Similarly, $e^{h'}$ gives $\mathcal{A}_2 = -i h'$ , using $\vec{dl} = \vec{e_2}$. These two conditions on $\vec{\mathcal{A}}$ mean that for $e^{h''}$ we must have:
\begin{equation}
\label{eq:condition_h''}
h'' = -h + h'
,
\end{equation}
using $\vec{dl} = \vec{e_2} - \vec{e_1}$. 

With the imaginary gauge field, the coupling matrices are then modified as $\tilde{H}^\dagger_{I \leftarrow J} = (t_1 e^{h'}, t_1 e^{h''}, \ldots, t_1 e^{h'}, t_1 e^{h''})$ and $\tilde{H}^\dagger_{J \rightarrow I} = (t_2 e^{-h'}, t_2 e^{-h''}, \ldots, t_2 e^{-h'}, t_2 e^{-h''})$.
The interference conditions now yield:
\begin{equation}
r_{0,1} = - \frac{t_1}{t_2} e^{2h}
,
\end{equation}
and
\begin{equation}
r_{0,2} = - \frac{t_1}{t_2} e^{2h'}
.
\end{equation}

Delocalization over the $e_1$-direction is achieved by requiring $|r_{0,1}| = 1$, namely choosing $h = h_0$.
Similarly, delocalization over the $e_2$-direction is realized with $h' = h_0$ so that $|r_{0,2}| = 1$.
One can notice that there is no further condition on $h''$ to delocalize the mode in the quasi-1D lattice. This is because of the vanishing amplitudes on the B and C sites.

Figure~\ref{fig:kagome_nh_prelim}(a),(b) show the normalized field profile $|\psi_{n,m}|$ of the zero-energy mode using $h = h_0$, $h' = h'' = 0$ and $h' = h_0$, $h = h'' = 0$, respectively. In Fig.~\ref{fig:kagome_nh_prelim}(a), the mode is localized on the bottom edge while being extended over the $e_1$-direction. On the other hand Fig.~\ref{fig:kagome_nh_prelim}(b) shows that the mode is localized on the left edge while being extended along the $e_2$-direction. 
It is worth noting that for the values of $h$, $h'$ and $h''$ chosen for drawing Fig.~\ref{fig:kagome_nh_prelim}, the spectrum has been changed compare to the Hermitian case.

However, provided Eq.~\ref{eq:condition_h''} holds, the introduction of the imaginary gauge field will only affect the localization property of the mode while keeping the spectrum unchanged. Therefore, in this case, the condition $|r_{0,i}| = 1$ does not correspond to band touching with the edge band and the bulk band~\cite{Kunst2017}.

Combining the two results obtained above for the delocalization of the zero-energy mode, we have $h = h' = h_0$ and, from Eq.~\ref{eq:condition_h''}, $h'' = 0$.
Figure~\ref{fig:kagome_nh}(a) shows that the numerically calculated spectrum of the kagome lattice in the rhombus geometry doesn't change when the imaginary gauge is introduced.
Figure~\ref{fig:kagome_nh}(c) plots the normalized field distribution of the zero-energy mode using $h = h' = h_0$ and $h'' = 0$. This topologically protected zero-energy mode is therefore extended over the whole bulk of the kagome lattice: it is a topological bulk mode in the 2D kagome lattice.

Generalization to higher dimensional lattices can be achieved using a similar procedure: starting from a topologically protected mode in lower dimension to delocalize this topological mode over that low dimensional bulk, and repeating this step with the higher dimension.

\section{Lasing in the non-Hermitian kagome lattice}
\label{section:kagome_nh_lasing}

The peculiarity of this extended topological mode is its vanishing amplitudes on the $B$ and $C$ sites but most of all, that it is topologically protected over a 2D bulk, and has a $\pi$-phase difference between non-vanishing sites.
This hints at the possibility to realize phase-locked broad-area topological lasers in 2D lattices.

\subsection{An active and non-Hermitian kagome lattice}

The passive design presented above can be extended to present a topological laser by using semiconductor ring resonators with gain $(g)$ and loss $(-g)$, and auxiliary rings for the imaginary gauge field~\cite{Longhi2018}.

The gain and loss configuration is chosen based on the SSH sublattices since we are looking for topological modes in the kagome lattice that originate from the topological mode in the SSH sublattice.
In the literature, it is well known that the SSH lattice in a $\mathcal{PT}$-symmetric configuration~\cite{Parto2018, Schomerus2013} (gain on site A, $E_A = i g$, and loss on site C, $E_C = -i g$) can preserve the topological protection of the zero-energy. Particularly, if the SSH lattice is in the unbroken $\mathcal{PT}$-symmetric phase, \ie $g < |t_2 - t_1|$, then the zero-energy mode is still guaranteed, from its pseudo-anti-Hermiticity~\cite{Parto2018, Takata2018}. 
When the energy is complex, we refer to the zero-energy as the real part of the energy being zero.

Since the only non-vanishing terms of the extended topological mode are located in the A sites, we expect lasing coming from these sites. Therefore we set gain on site A, $E_A = i g$, and lossy rings on the B and C sites, $(E_B = E_C = -i g)$, as shown in Fig.~\ref{fig:quasi_1D_kagome}(b). 

\begin{figure}
\center
\includegraphics[width=\columnwidth]{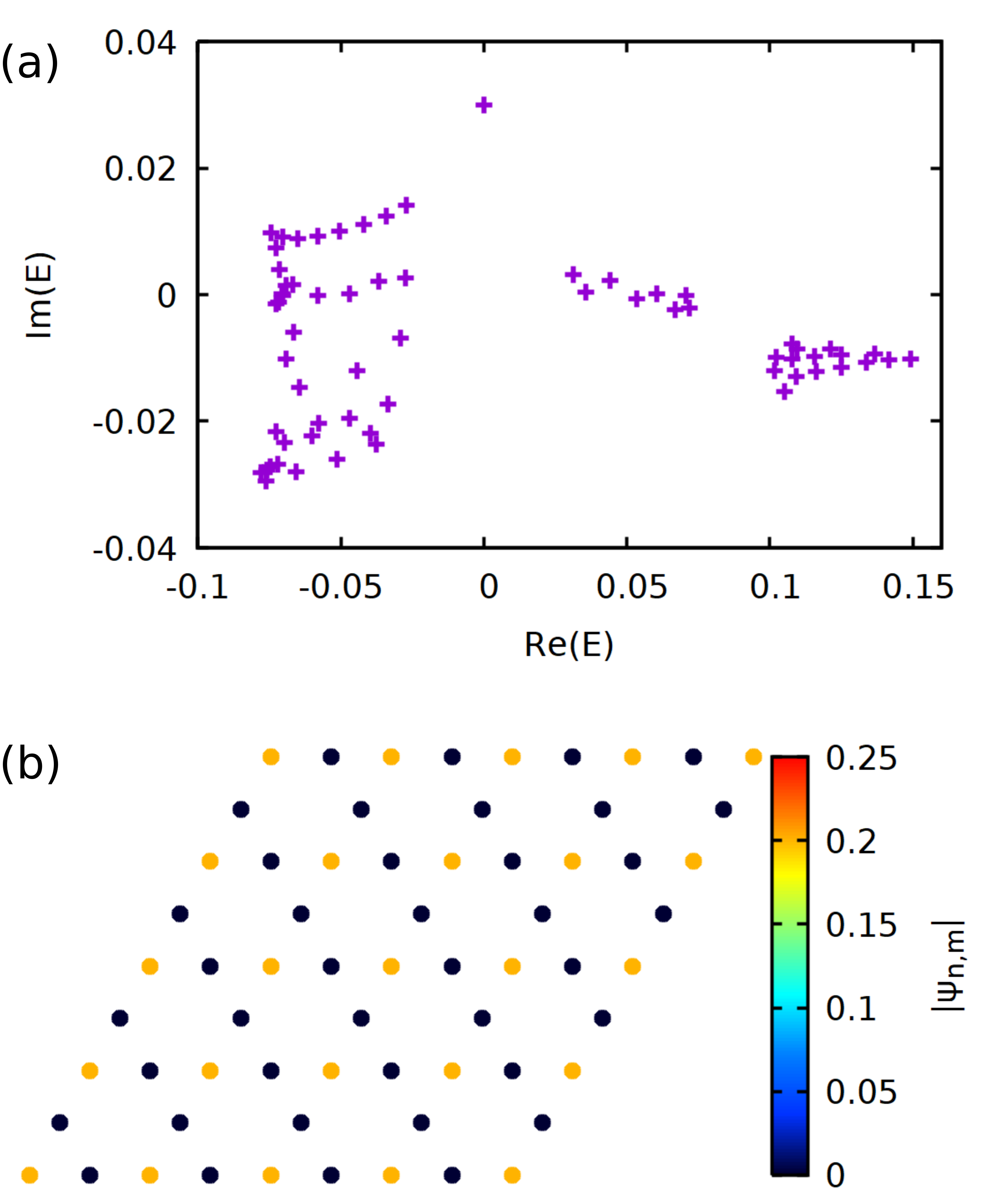}
\caption{
(a) Spectrum of the active kagome lattice in the rhombus geometry.
(b) Normalized field profile $|\psi_{n,m}|$ of the zero-energy mode.
We have $N_s = 11$ sites both in the $I$ lattice and the quasi-1D lattice, $t_1 = 0.02$, $t_2 = 0.06$, $g = 0.03$ and $h = h' = h_0$, $h'' = 0$.
}
\label{fig:kagome_nh_active}
\end{figure}

Figure~\ref{fig:kagome_nh_active}(a) shows the spectrum in the complex plane, numerically calculated with $E_A = i g$, $E_B = E_C = -i g$, $h = h' = h_0$ and $h'' = 0$. This demonstrates that the zero mode is present. Because of the gain on the A sites, \ie where the zero-energy mode is non-vanishing, the zero-energy mode has higher gain compared to other modes. 
Figure~\ref{fig:kagome_nh_active}(b) displays the normalized field profile $|\psi_{n,m}|$ of the zero-energy mode. This shows that, although active design has been considered, the delocalization of the zero-energy mode is not altered.

\subsection{Temporal dynamics of the zero-energy mode}

In the frequency analysis, we have seen it is possible to have an active non-Hermitian kagome lattice with an extended topological mode. It is now interesting to see whether this mode presents temporal instabilities.

The previous analysis provides a simple physical model of the active non-Hermitian kagome lattice. 
It has been shown that temporal instabilities in the laser array may prevent phase-locking and reduce the laser quality or even dominate and suppress the topological signature of the corresponding lasing mode~\cite{Longhi2018, Longhi2018a}. 
Therefore time domain modelling of the mode dynamics is essential for determining whether lasing is stable.

We consider the laser rate equation for modelling the gain in the active rings A~\cite{Winful1990, Li1992, Longhi2018}. Because the zero-energy mode has vanishing amplitude on the ring resonators B and C, linear loss is chosen for those rings.
The laser rate equation in the kagome lattice, with $h = h' = h_0$ and $h'' = 0$, is then:
\begin{multline}
i \frac{d a_{n,m}}{dt} = 
\frac{1}{\tau_p}(1 - i \alpha) Z_{n,m} a_{n,m} + t_1 e^{-h_0} b_{n,m} + t_1 e^{-h_0} c_{n,m} \\ 
+ t_2 e^{h_0} b_{n,m-1} + t_2 e^{h_0} c_{n-1,m}
,
\end{multline}
\begin{multline}
i \frac{d b_{n,m}}{dt} = 
-i g_B b_{n,m} + t_1 e^{h_0} a_{n,m} + t_1 c_{n,m} \\ 
+ t_2 e^{-h_0} a_{n,m-1} + t_2 c_{n-1,m+1}
,
\end{multline}
\begin{multline}
i \frac{d c_{n,m}}{dt} = 
-i g_C c_{n,m} + t_1 e^{h_0} a_{n,m} + t_1 b_{n,m} \\ 
+ t_2 e^{-h_0} a_{n+1,m} + t_2 b_{n+1,m-1}
,
\end{multline}
\begin{multline}
\tau_s \frac{d Z_{n,m}}{dt} = p_A - Z_{n,m} - (1 + 2 Z_{n,m}) |a_{n,m}|^2
,
\end{multline}
where $a_{n,m}$, $b_{n,m}$ and $c_{n,m}$ are the modal amplitudes on the site A, B and C and in the $(n,m)$-th unit cell, respectively. $n$ and $m$ stand for the unit-cell index in the SSH lattice and quasi-1D lattice, respectively. $Z_{n,m}$ is the normalized excess carrier density in the active ring A, $\tau_p$ and $\tau_s$ are the photon and spontaneous carrier lifetimes, respectively, $\alpha$ the linewidth enhancement factor, $p_A$ the normalized excess pump current in the ring A, and $g_B$ and $g_C$ the linear loss in the ring B and C, respectively.

\begin{figure*}
\center
\includegraphics[width=2\columnwidth]{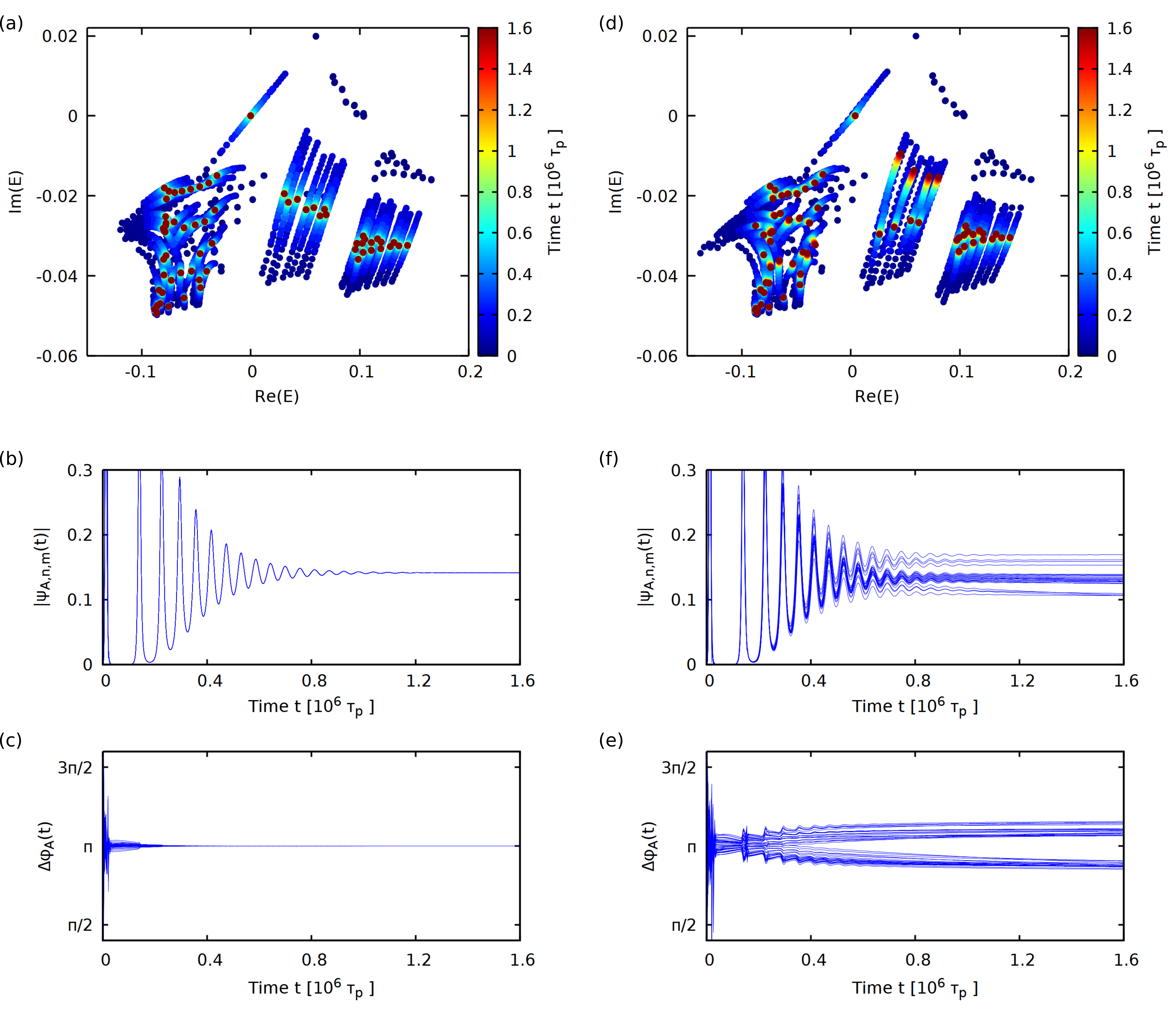}
\caption{
Time evolution of 
(a) the instantaneous spectrum of the kagome lattice in rhombus geometry 
(b) the amplitudes of all the $A$ sites, and 
(c) the phase differences between the adjacent $A$ sites when there is no disorder.
Similarly for (d), (e) and (f) when asymmetric disorders is introduced.
The parameters used are: $\tau_p = 40 \text{ps}$, $\alpha = 3$, $\tau_s = 80 \text{ns}$, $p_A = 0.02$, $g_B \tau_p = g_C \tau_p = 0.05$, $\delta_B \tau_p = \delta_C \tau_p = 0.01$, $t_1 \tau_p = 0.02$, $t_2 \tau_p = 0.06$.
}
\label{fig:time_evolution}
\end{figure*}

The coupled-mode equations are integrated using random noise of field amplitudes between $[0, 0.01]$ and equilibrium carrier density $Z_{n,m} = p_A$ as initial condition.
The random noise as initial condition is chosen to trigger non-linear behaviour and see whether or not the mode is stable. 
The parameters are chosen similar to Ref.~\cite{Winful1990, Li1992, Longhi2018} and are typical for semiconductor lasers.
Figure~\ref{fig:time_evolution}(a) displays the time evolution of the instantaneous spectrum of the kagome lattice in rhombus geometry. It shows that after a transient time, the system reaches a single laser mode regime. The laser mode is the topological zero-energy mode with $\text{Im}(E) = 0$.
Figure~\ref{fig:time_evolution}(b) and (c) display the time series of the field amplitudes at all the A sites and the adjacent phase difference between the A rings.
This shows that after a transient regime, only the zero-energy mode survives and reaches a steady state. 
The amplitudes of all the $A$ sites are equally distributed over the bulk and have a fixed $\pi$-phase difference.
The laser system obtained is therefore broad-area and phase-locked.

In addition to this interesting broad-area and phase-locked feature, the laser mode is topologically protected.
Figure~\ref{fig:time_evolution}(d) shows the spectrum of the system when an asymmetric perturbation on the couplings, $\delta t_{1\pm}$, is added: $t_1 e^{h} \rightarrow (t_1 + \delta t_{1+}) e^{h}$ and $t_1 e^{-h} \rightarrow (t_1 + \delta t_{1-}) e^{-h}$. This asymmetric perturbation accounts for perturbation in the coupling strength as well as for the imaginary gauge field. One can see that the zero-energy mode is still present.
However one can see in Fig.~\ref{fig:time_evolution}(e) and (f) a slightly different behaviour in the time series of the field amplitudes and the phase difference between the A rings. They do not reach the same values in amplitudes and phase differences.
The amplitudes are not equally distributed but have slightly different offsets because of the different couplings between each sites: a single choice for the imaginary gauge field $h$ and $h'$ cannot satisfy the $|r_{0,i}|=1$ conditions between each sites.
Similarly, the phase differences are no longer equal to $\pi$ due to the non-linear Henry factor $\alpha$.
Nevertheless, here, the addition of perturbation in the coupling strengths does not give rise to unstable behaviour in the amplitudes and phases of the topological extended mode.

\begin{figure*}
\center
\includegraphics[width=1.7\columnwidth]{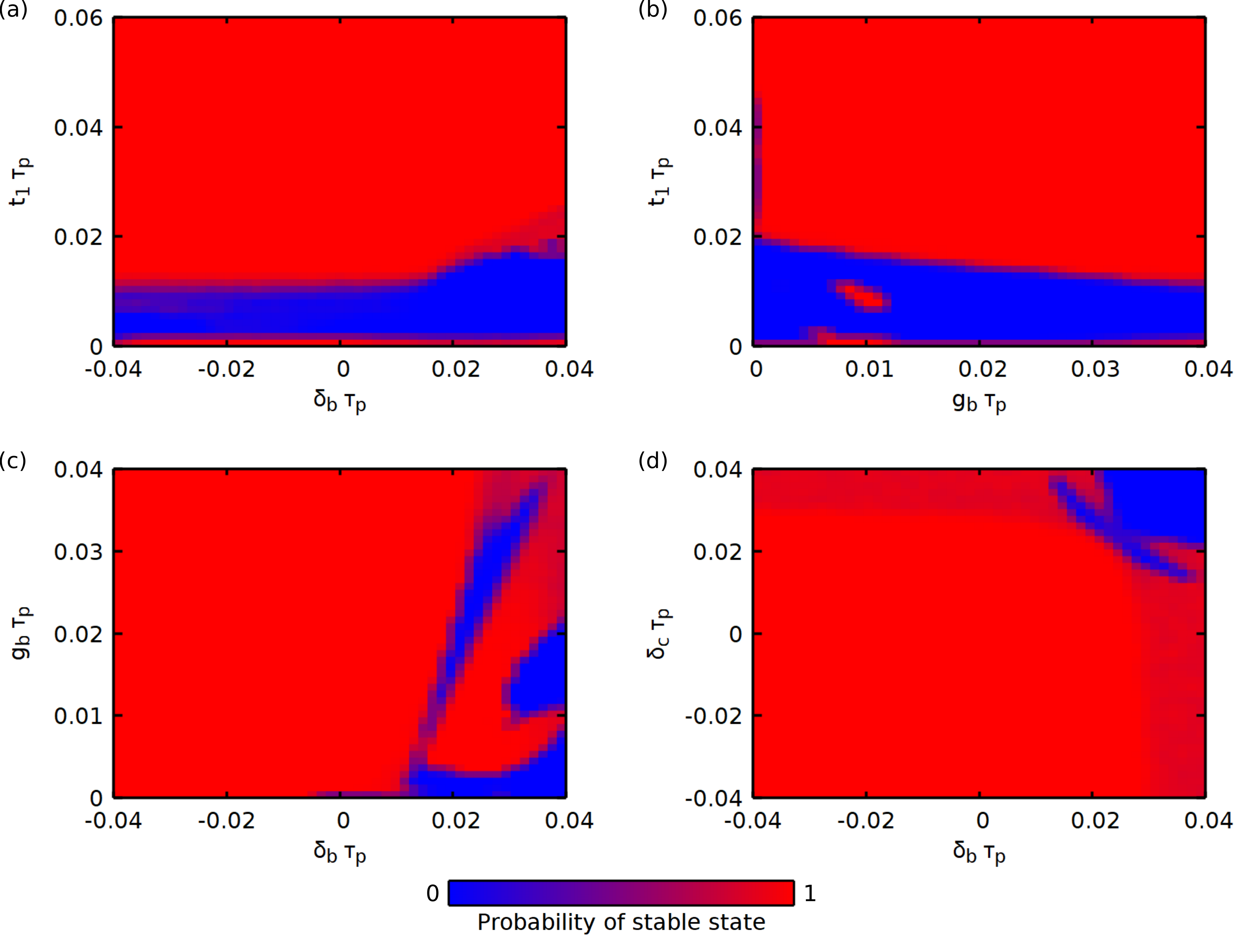}
\caption{
Stability diagrams of the topological extended mode. 
The color plot corresponds to the probability of the system being in the stable regime over 200 realizations in the random initial conditions of the mode amplitudes.
The fixed parameters are the same in Fig.~\ref{fig:time_evolution}.
}
\label{fig:stability_diagram}
\end{figure*}

With the parameters chosen in Fig.~\ref{fig:time_evolution}, the system reaches a stationary state which does not evolve into random oscillation in their amplitudes and phase differences. This means that the zero-energy mode does not suffer from non-linear instabilities. 
Even though the spatial stability of the topological mode is guaranteed by its topological invariant, it is worth looking at its temporal stability in the parameter space to delimit the region where temporal instabilities arise due to the non-linear terms.
In the following, we will refer to the stable regime, the regime of single mode lasing in the topological extended mode. Therefore, we say the system to be unstable (or stable) if oscillations are present (or not) in their amplitudes or phase differences.

Figure~\ref{fig:stability_diagram} shows the stability diagram of the topological extended laser mode for different slices of the parameter space. These demonstrate stable phase-locking of the non-Hermitian gauge laser array in a relatively large region of the parameter space.
Numerical results show the stability of the topological extended mode requires a minimum coupling strength (Fig.~\ref{fig:stability_diagram}(a)-(b)).
One reason is that the $\mathcal{PT}$-symmetric phase is broken when the couplings are too small compared to the gain and loss~\cite{Parto2018}: the system is no longer in the single lasing mode regime.
The second reason is that the non-reciprocal dissipative couplings need to be high enough in order to reach a (soft) synchronization~\cite{Ding2019}.
On the other hand, instabilities arise when the detunings $\delta_b$ and $\delta_c$ are too high (Fig.~\ref{fig:stability_diagram}(c)). The major critical case is with positive detunings (Fig.~\ref{fig:stability_diagram}(c)-(d)).
However, stability is retrieved if the dissipation is large enough to compensate the detunings in the rings B and C.

\begin{figure}
\center
\includegraphics[width=\columnwidth]{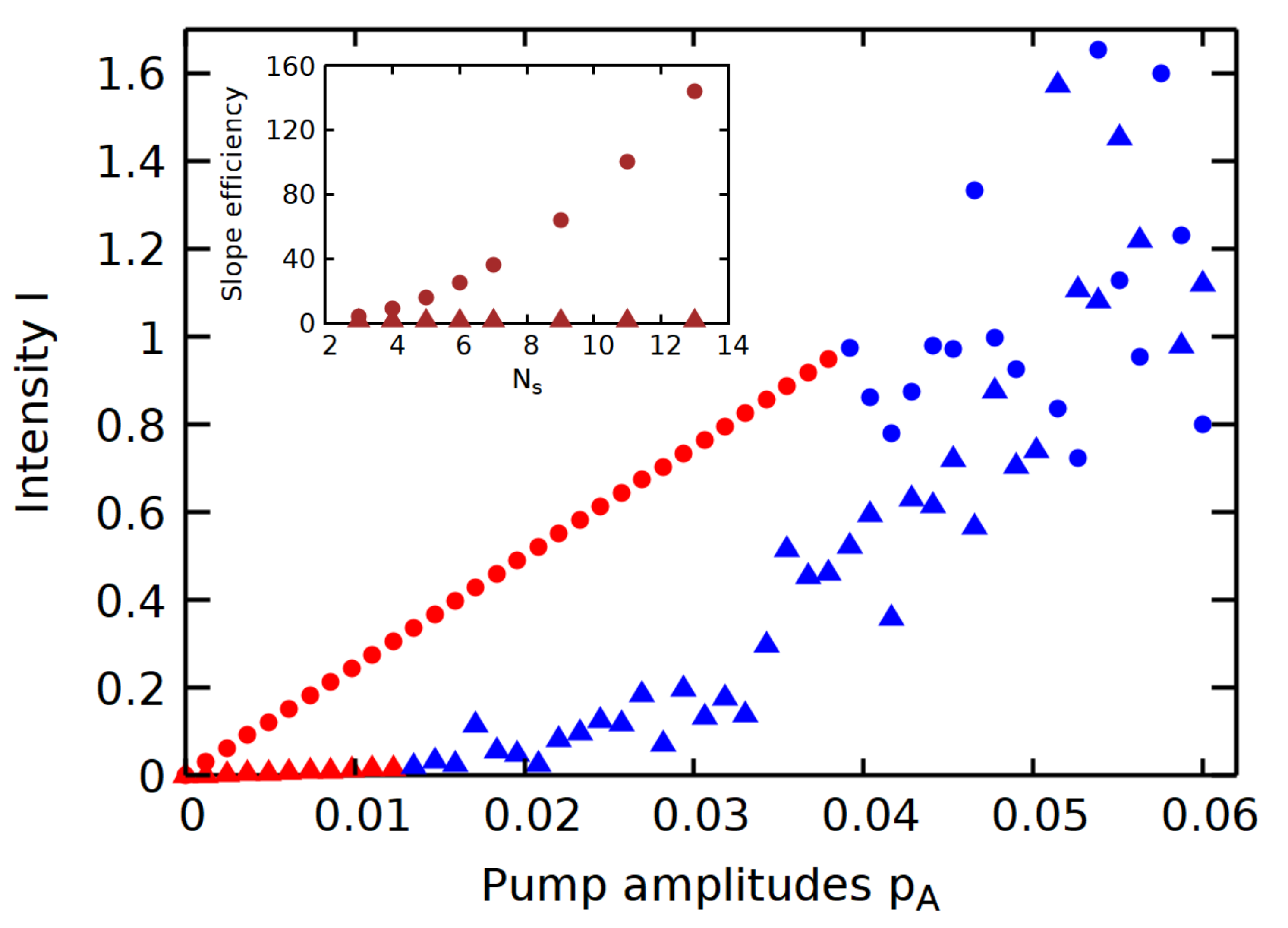}
\caption{
Total intensity against pump amplitudes $p_A$ for the localized and delocalized topological mode with $N_s = 11$. 
The triangles (circles) correspond to the topological edge (bulk) laser.
The colors plot indicate if the system in the unstable (blue) or stable (red) regime.
The inset shows the dependency of the slope efficiency with respect to $N_s$.
The parameters are the same as in Fig.~\ref{fig:time_evolution}.
}
\label{fig:slope_efficiency}
\end{figure}

Extended lasing modes present an important advantage in getting a better slope efficiency, compared to the compact lasing mode.
Figure~\ref{fig:slope_efficiency} plots the total intensity, $I = \sum_{m,n} |\psi_{n,m}|^2$ against the pump amplitudes $p_A$ for the compact and extended lasing mode where the color plot corresponds to the system being in the stable or unstable regime. 
The numerical results show that the compact mode has lower slope efficiency compared to the extended mode.
The remarkable difference is the scaling of the slope efficiency with the size of the system for the delocalized mode while it remains constant for the compact mode, as shown in the inset of Fig.~\ref{fig:slope_efficiency}.
This is because of the extended nature of the mode whose contribution to the lasing intensity increases with the size of the system.

\begin{figure*}
\center
\includegraphics[width=1.7\columnwidth]{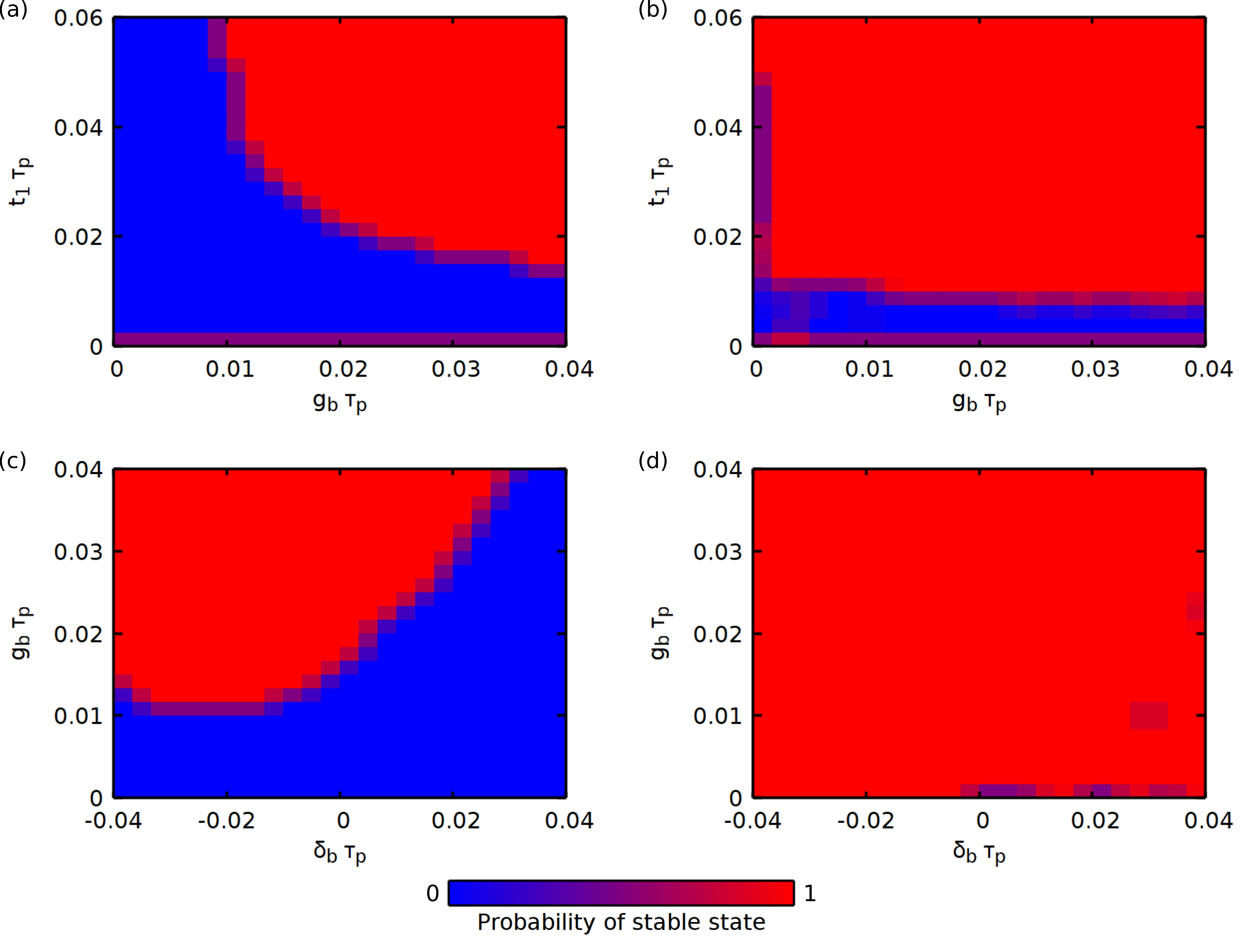}
\caption{
Stability diagrams of the topological (a)-(c) compact mode, (b)-(d) extended mode. 
The color plot corresponds to the probability of the system being in the stable regime. 
The fixed parameters are the same in Fig.~\ref{fig:time_evolution} except here $p_A = 0.01$.
}
\label{fig:stability_diagram_comparison}
\end{figure*}

Clearly, the imaginary gauge field helps to stabilise the system in the zero-energy lasing mode.
The extended nature of the topological mode over the bulk allows the zero-energy mode to carry all the gain of the system while suppressing all the other bulk modes.
The color plot in Fig.~\ref{fig:slope_efficiency} indicates that the compact mode reaches an unstable regime for relatively low values of pump intensity whereas the extended mode is unstable for higher pump intensities.
The stability diagrams for the compact mode are shown in Fig.~\ref{fig:stability_diagram_comparison}(a)-(c).
Compared to the extended mode in Fig.~\ref{fig:stability_diagram_comparison}(b)-(d), these diagrams demonstrate that extended modes have, indeed, larger regions of stability in the parameter space than compact modes.

\begin{figure}
\center
\includegraphics[width=\columnwidth]{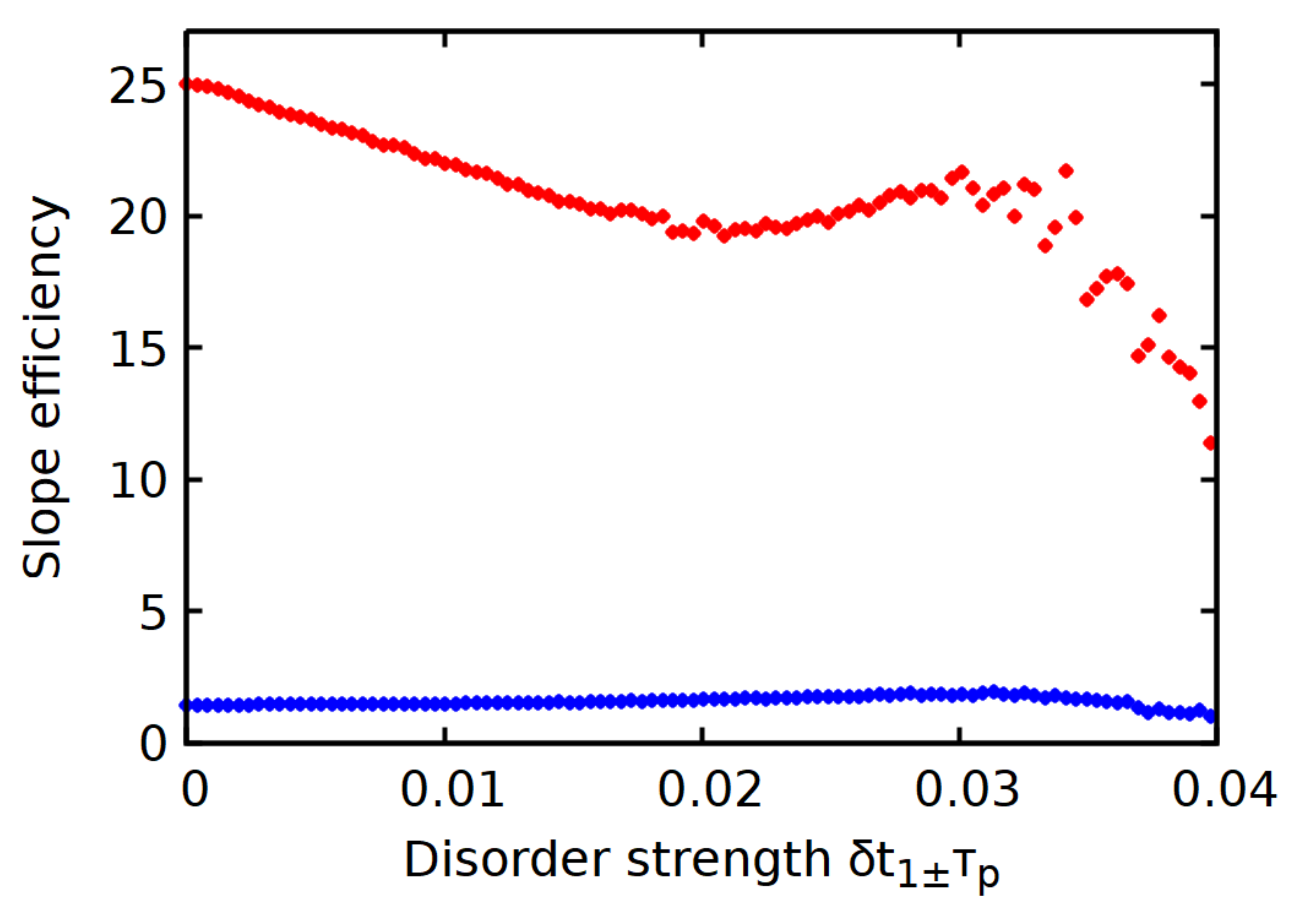}
\caption{
Slope efficiency against disorder strength $\delta t_{1\pm}$ (for $N_s = 11$). The blue (red) dots correspond to the topological edge (extended) laser.
For each disorder strength, the slope efficiency is average over 800 random realizations of the disorder.
The parameters are the same as in Fig.~\ref{fig:time_evolution}.
}
\label{fig:slope_vs_disorder}
\end{figure}

Figure~\ref{fig:slope_vs_disorder} demonstrates the slope efficiency is robust against asymmetric disorder in the couplings, for both the compact and extended mode, as expected the topological nature of the lasing mode.
The compact mode has a constant slope efficiency with increasing disorder strength.
While the extended mode gives varying slope efficiency because of varying field distribution of the extended mode in the bulk as explain before, its slope efficiency remains higher compared to the compact mode.
However, for high values of disorder strength in the couplings, the slope efficiency of the extended mode starts to decrease.
This is explained because the delocalization of the topological mode, originating from non-reciprocal couplings, is highly perturbed by the asymmetric perturbation $\delta t_{1\pm}$ in the couplings: the topological mode may not be completely delocalized anymore.

\section{Conclusion}
To summarize, we have shown a procedure to get topological modes extended over a $d$-dimensional bulk using an imaginary gauge field. 
In particular, we have demonstrated the existence of a topological extended mode in the kagome lattice in the rhombus geometry.
This topological extended mode in the kagome lattice has been studied in the context of a lasing system where the laser rate equation is included to take into account non-linear effect.
Investigating its temporal stability, we proved that stable topological broad-area phase-locked laser operation is possible in a large region in the parameter space.

Furthermore, it has been shown that the topological extended mode presents clear advantages over the topological localized mode.
The extended nature of the former topological mode over the bulk enhances its temporal stability, and yields higher slope efficiency that scale with the size of the system.
In terms of footprint, a higher dimensional extended mode is also more advantageous compared to their lower dimensional one since at equivalent slope efficiency, for example, the system occupies a smaller  region in real space when increasing the dimensionality of the system.
This can lead to applications where transport of high energy density is possible.

\begin{acknowledgments}
This research was undertaken using the supercomputing facilities at Cardiff University operated by Advanced Research Computing at Cardiff (ARCCA) on behalf of the Cardiff Supercomputing Facility and the HPC Wales and Supercomputing Wales (SCW) projects. We acknowledge the support of the SCW projects and Sêr Cymru II Rising Star Fellowship (80762-CU145 (East)), which are part-funded by the European Regional Development Fund (ERDF) via the Welsh Government. 
\end{acknowledgments}


\bibliographystyle{apsrev4-1}
\bibliography{ref}

\end{document}